\documentclass[pageno]{jpaper}

\usepackage[normalem]{ulem}

\usepackage{enumitem}
\usepackage{listings}          % format code
\usepackage{booktabs}
\usepackage{amsmath}
\usepackage{xspace}
\usepackage{xcolor,colortbl}
\usepackage{array}
\usepackage{subfig}
\usepackage[nohyperlinks, printonlyused, withpage, smaller, nolist]{acronym}
\usepackage{balance}

\definecolor{Gray}{gray}{0.85}

\newcommand{\argmax}[1]{\underset{#1}{\operatorname{arg}\,\operatorname{max}}\;}

\newcommand{\TSPostComment}[1]{}

\newcommand{\ourparagraph}[1]{\smallskip\noindent\emph{#1}}

\newcommand{\Bus}{\mathsf{bus}}
\newcommand{\Cache}{\mathsf{cache}}
\newcommand{\RAM}{\mathsf{RAM}}
\newcommand{\slowdown}{\mathsf{slowdown}}
\newcommand{\tuned}{\mathsf{tuned}}
\newcommand{\resourcetunedenemies}{\mathsf{ResourceTunedEnemies}}
\newcommand{\rankedenvironments}{\mathsf{RankedEnvironments}}

\newcommand{\PiThree}{Pi3\xspace}
\newcommand{\Dragon}{410c\xspace}
\newcommand{\IntelJoule}{570X\xspace}
\newcommand{\Nano}{T3\xspace}
\newcommand{\Banana}{M3\xspace}

\newcommand*{\aboverulesepcolor}[1]{%
  \noalign{%
    \begingroup
      \color{#1}%
      \hrule height\aboverulesep
    \endgroup
    \kern-\aboverulesep
  }%
}

\begin{document}

\title{
  %An auto-tuning framework for multi-core interference analysis
  %Caring When Sharing: Multi-core Interference Tuning and Analysis
  Do Your Cores Play Nicely?\\ A Portable Framework for Multi-core Interference Tuning and Analysis
}

\author{
  Dan Iorga\\
  Imperial College London\\
  %\texttt{d.iorga17@imperial.ac.uk}
  d.iorga17@imperial.ac.uk
  \and
  Tyler Sorensen\\
  Princeton University\\
  %\texttt{ts20@cs.princeton.edu}
  ts20@cs.princeton.edu
  \and
  Alastair F. Donaldson\\
  Imperial College London\\
  %\texttt{afd@imperial.ac.uk}
  alastair.donaldson@imperial.ac.uk
}

\begin{acronym}
\acro{RAN}{Random search}
\acro{SA}{Simulated Annealing}
\acro{BO}{Bayesian Optimisation}
\end{acronym}

\date{}
\maketitle

\thispagestyle{empty}

\lstset{columns=fullflexible}
\lstset{
  language=C, % choose the language of the code
  showspaces=false, % show spaces adding particular underscores
  showstringspaces=false, % underline spaces within strings
  showtabs=false, % show tabs within strings adding particular underscores
  tabsize=8, % sets default tabsize to 2 spaces
  captionpos=b, % sets the caption-position to bottom
  mathescape=true, % activates special behaviour of the dollar sign
  basicstyle=\scriptsize\tt,
  columns=fullfelxible,
  xleftmargin=2em,
  numbers=left,
  %commentstyle=\rmfamily\itshape,
  morekeywords={barrier,kernel,global,local,__device__,__syncthreads,__global__,bool,then},
  escapeinside={(*@}{@*)}
}

% subfig doesn't like lstlistings, so put them in a box, as recommended at:
% https://tex.stackexchange.com/questions/34946/how-can-i-put-lstlisting-block-into-subfloat-block
\newsavebox{\firstlisting}
\begin{lrbox}{\firstlisting}% Store first listing
      \begin{lstlisting}[numbers=left, xleftmargin=20em, basicstyle=\footnotesize\ttfamily]
void vec_add(float *A, float *B, float *C, int SIZE) {
  for (int i = 0; i < SIZE; i++)
    C[i] = A[i] + B[i];
}
\end{lstlisting}
\end{lrbox}

\newsavebox{\secondlisting}
\begin{lrbox}{\secondlisting}% Store first listing
\begin{lstlisting}[numbers=left, xleftmargin=20em, basicstyle=\footnotesize\ttfamily]
void cache_enemy(byte* scratch) {
  while(1)
    for (int i = 0: i+=STRIDE; i < BUFFER_SIZE)
      ACCESS(&(scratch[i]));
}
\end{lstlisting}
\end{lrbox}

\begin{abstract}

Multi-core architectures can be leveraged to allow independent processes to run in parallel. However, due to resources shared across cores, such as caches, distinct processes may interfere with one another, e.g.\ affecting execution time. Analysing the extent of this interference is difficult due to: (1)~the diversity of modern architectures, which may contain different implementations of shared resources, and (2)~the complex nature of modern processors, in which interference might arise due to subtle interactions.  To address this, we propose a black-box auto-tuning approach that searches for processes that are effective at causing slowdowns for a program when executed in parallel.  Such slowdowns provide lower bounds on worst-case execution time; an important metric in systems with real-time constraints.

Our approach considers a set of parameterised ``enemy'' processes and ``victim'' programs, each targeting a shared resource. The autotuner searches for enemy process parameters that are effective at causing slowdowns in the victim programs. The idea is that victim programs behave as a proxy for shared resource usage of arbitrary programs. We evaluate our approach on: 5 different chips; 3 resources (cache, memory bus, and main memory); and consider several search strategies and slowdown metrics. Using enemy processes tuned per chip, we evaluate the slowdowns on the autobench and coremark benchmark suites and show that our method is able to achieve slowdowns in $98\%$ of benchmark/chip combinations and provide similar results to manually written enemy processes.

\end{abstract}

\section{Introduction}

Multi-core processors have seen widespread adoption, with nearly every consumer device containing more than one independent processing unit.  However, due to shared resources and their corresponding arbitration mechanisms, e.g.\ cache hierarchies and protocols, reasoning about program behaviours on multi-core processors can be significantly more complex than on their single-core predecessors. Interference can affect non-functional properties of otherwise entirely separate processes, e.g.\ the execution time of a program on a multi-core processor can vary greatly depending on shared resource contention. Because of these issues, the \emph{Worst Case Execution Time} (WCET) of an application on a multi-core chip is difficult to derive, and as a result, multi-core processors remain challenging to deploy in systems with hard or soft real-time constraints.

% Mitigation is not fully possible
Because of this, prior work has identified \emph{interference paths}, where the contention and arbitration of shared resources might impact program execution time~\cite{5347560,5347562,5347561,Berthon2016,Brindejonc2014}. Components of interference paths include caches, memory buses, and main memory systems. Although not widely adopted, various schemes have been proposed to limit this interference. Such schemes require either hardware support, e.g.\ cache partitioning~\cite{6375555}, or invasive software modifications, e.g.\ bank-aware memory allocation and bandwidth reservation~\cite{6531079,6925999}. However, even with these schemes, interference can still be substantial~\cite{7461361}. As a result, there is immediate and pragmatic interest in detecting and quantifying interference effects rather than aiming to mitigate them entirely.

% It is not just our idea, others are interested too
In this vein, various techniques have been investigated to quantify the effects of interference on real-time properties, and evaluated for specific multi-core architectures~\cite{Radojkovic:2012:EIS:2086696.2086713, 6214768, Fernandez:2012:ASN:2380356.2380389}. Typically, work in this domain consists of: manually developing small programs designed exclusively to stress an interference path (called \emph{enemy processes} in this work), executing a set of enemy processes on all but one core of a multi-core system (called a \emph{hostile environment} in this work), and evaluating the execution time of a sequential \emph{Software Under Test} (SUT) on the remaining core. The slowdown observed in the SUT from the hostile environment is a useful quantification of interference effects.

This prior work requires the manual design of hostile environments and corresponding enemy processes, presenting two immediate limitations. First, manual enemy process design is not \emph{portable} across architectures, as different architectures may have different implementations, or their shared resources may be configured differently. Thus, manual effort is required for each new target architecture. Second, hand-tuned enemy processes may not be as effective as possible, due to subtle interactions that are difficult to derive from available documentation, and thus unlikely to be considered by human designers.

\begin{figure}
  \subfloat[A vector addition program that computes $C \leftarrow A + B$]{ \hspace{1em} \usebox{\firstlisting}\label{fig:example-vec-add}} \\
  \subfloat[An enemy template that targets the cache by accessing a region of memory in cache line sized strides; values in CAPS are parameters]{ \hspace{1em} \usebox{\secondlisting}\hspace{1.6cm}\label{fig:example-enemy}}
\caption{Example of an SUT~(a) and an enemy process~(b)\label{fig:intro-example}}
\end{figure}

In this work, we aim to address both limitations. The heart of our contribution is an auto-tuning method that can tune enemy process parameters to be effective at slowing down an SUT. For each interference path considered, our approach takes a parameterised enemy process (called an \emph{enemy template}) and a corresponding \emph{victim program}, which is designed to be especially vulnerable to the particular interference path. We then employ auto-tuning to search for enemy process parameters that are effective at causing a slowdown in the corresponding victim program. After obtaining tuned enemy processes, we employ a second level of tuning over all victim programs to obtain a combination of enemy processes, which can be deployed as a hostile environment. The tuning approach can be run on many chips using the same enemy templates and victim programs to produce chip-specialised hostile environments.

We illustrate the problems with manually-designed enemy processes, and explain at a high-level how our auto-tuning approach overcomes these limitations, using an example.

\paragraph{Example of Portability Limitation}
%
% \TSComment{The size of the vector is never discussed. I think it should be hard-coded into the program text and mentioned here that the size is too big to fit in a local cache}

\begin{table}
  \footnotesize
  \centering
      \caption{Parameters for the enemy template of
        Figure~\ref{fig:example-enemy} for the \PiThree and \IntelJoule
        boards: {\tt BUFFER\_SIZE} is given in KB; {\tt STRIDE} is given in
        bytes; and {\tt ACCESS} is given by a sequence of
        store~(S) and load~(L) operations}
      \label{tab:example-parameters}
  \begin{tabular}{l r r r r }
    \toprule
    &  \multicolumn{2}{c}{\PiThree} & \multicolumn{2}{c}{\IntelJoule} \\
    &\multicolumn{1}{c}{hand-tuned} & \multicolumn{1}{c}{auto-tuned} & \multicolumn{1}{c}{hand-tuned} & \multicolumn{1}{c}{auto-tuned} \\
    \midrule
        {\tt BUFFER\_SIZE} & 512 & 20480 & 2048 & 40960  \\
        {\tt STRIDE} & 64 & 262 & 64 & 40960 \\
        {\tt ACCESS} & SL & SLSSL & SL & SS \\
    \bottomrule
    \end{tabular}

\end{table}
\begin{figure}
  \centering
  \includegraphics[width=.49\linewidth]{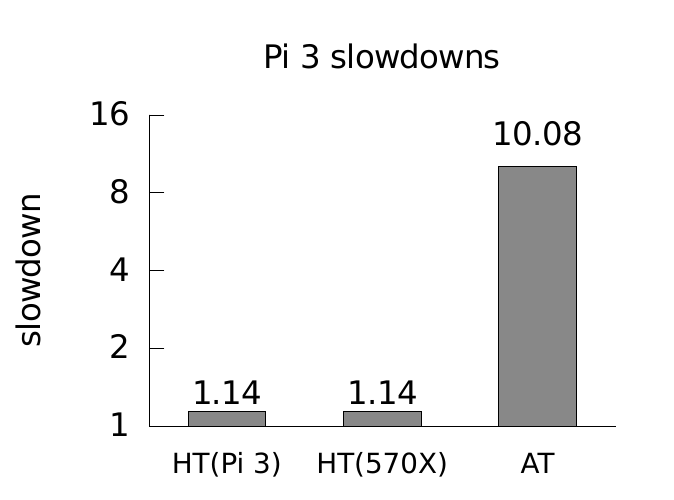}
  \includegraphics[width=.49\linewidth]{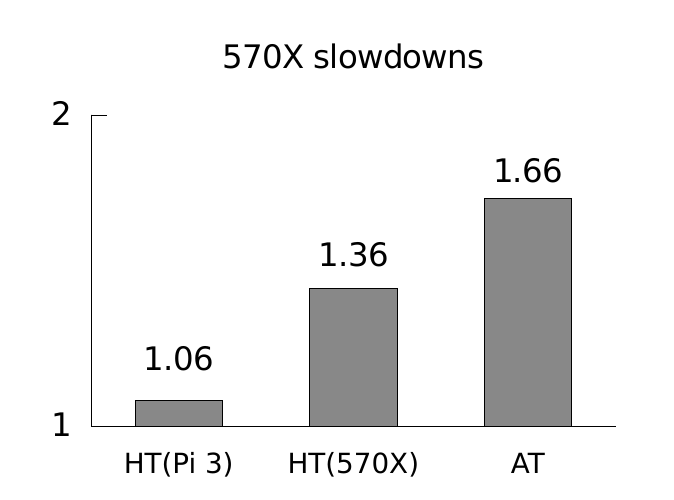}
  \caption{Slowdowns caused by different hostile environments on the
    \PiThree and \IntelJoule: HT denotes enemy processes hand-tuned
    for the chip given in (), and AT denotes auto-tuned enemy
    processes for the target chip }
 \label{fig:example-slowdowns}
\end{figure}

Figure~\ref{fig:example-vec-add} shows a simple vector addition
program. The {\tt SIZE} parameter can be set to a value (in this case
we use 16K) such that the vector data is not able to fit in a
entirely in a core-local cache; thus, memory accesses must go through
shared caches at some point. Suppose we want to assess
the potential interference between this program (the SUT) running on
one core and a set of independent programs running on other cores.
The program of Figure~\ref{fig:example-enemy} is an example of an
enemy process: the sole task of the program is to exercise an
interference path. This particular enemy process is designed to cause
cache contention by looping over a memory region the size of the
shared cache and accessing memory at a stride of the cache line
size. To explore potential interference, the execution time of the SUT
executing on one core can be measured while multiple instances of the
cache enemy processes are running on other cores in the system.

As the enemy process code of Figure~\ref{fig:example-enemy} shows,
there are three parameters that need to be instantiated: (1)~size of
the shared cache; (2)~cache line size; and (3)~memory instructions,
i.e.\ loads or stores, used to access memory. To provide suitable
values, the target processor must be known. For example, on a 4-core
Raspberry Pi 3 B (abbreviated to \PiThree), the shared cache size is
512KB, the cache line is 64 bytes and we use a single store followed
by a load as the instructions. The execution time of the SUT shows a
$1.14\times$ slowdown when executed in parallel with this enemy
processes running on the three additional cores.

Now, if we consider a different processor, say a
4-core Intel Joule 570X (abbreviated to \IntelJoule), we might try the
same experiment using the same parameters from the \PiThree. In this case,
we observe a slowdown of $1.06\times$. We then might try a different
enemy process, tuned to the architectural details of the
\IntelJoule. This changes the size of the shared cache to 2048KB but
keeps the same cache line size. With this new enemy process, a much greater
slowdown of $1.36\times$ is observed. However, when the \IntelJoule
enemy process is used on the \PiThree, the slowdown observed is
$1.14\times$, exactly the same as the original \PiThree enemy
process. Since the \PiThree and the \IntelJoule have the same same cache line size and the same associativity but different cache sizes, the enemy processes will access the cache in a simillar manner but the \PiThree enemy process will not access the entire \IntelJoule cache.

These results are intuitive: enemy processes tuned to a particular architecture will be less effective when running on a different architecture. Prior work in this area has largely focused on such enemy processes, i.e.\ hand-tuned for a particular architecture.  In the case where enemy processes are written at the assembly level, the situation is even worse: enemy processes designed with respect to one ISA are inapplicable to a processor with a different ISA.

\paragraph{Example of Effectiveness Limitation}
Recall that the hand-tuned enemy processes showed an SUT slowdown of
$1.14\times$ and $1.36\times$ for the \PiThree and \IntelJoule,
respectively.  These slowdowns were achieved by making reasonable
human judgements for enemy process parameters.  If instead, using the
methodology described in the remainder of this paper, the enemy
template is \emph{auto-tuned} for two hours, different parameters can
be found (summarised in Table~\ref{tab:example-parameters}).

The values found by auto-tuning do not correspond to any architectural features that we are aware of, and it seems unlikely that a human designer with detailed knowledge of these processors would guess them. However, the slowdowns of the SUT using this auto-tuned hostile environment increase to $10.08\times$\footnote{This slowdown is alarming, but we have rigorously validated this result and see similar values for the \PiThree throughout this work} and $1.66\times$ for the \PiThree and \IntelJoule, respectively. The slowdown results are summarised in Figure~\ref{fig:example-slowdowns}. Thus, an auto-tuning methodology for enemy processes (1)~requires less architectural knowledge and manual effort than hand-tuned approaches of previous work, and (2)~can provide greater slowdowns than reasonable hand-tuned enemy processes.

\paragraph{Contributions}

We present an auto-tuning methodology for hostile environments and the corresponding enemy processes. This approach can be employed for different chips, removing the need for the detailed architectural expertise required by prior works in this area. Additionally, because of the many configurations explored by the tuning approach, our methods may discover non-intuitive parameters that cause slowdowns beyond what might be developed through hand-optimisation, as illustrated in the above example.

%The auto-tuning approach takes a set of enemy templates: parameterised programs each of which target an abstract interference path. For each enemy template, a victim program is also required. The victim program is a microbenchmark that aims to make heavy use of an interference path. We present a two-phase tuning process: first, each enemy process, victim program pair is tuned; second, the combinations of tuned enemy processes are tuned to be effective across all victim programs. The combination of tuned enemy processes can be deployed as a hostile environment for doing timing analysis on an SUT.

We illustrate our approach by creating a hostile environment on five
different chips for three interference paths: cache, memory bus, and
the main memory system. This requires three enemy process, victim
program pairs. For tuning, we explore three search strategies (random search,
simulated annealing and Bayesian optimisation), and report on the
effectiveness of each.

Finally, we assess the effectiveness of our approach at causing slowdowns in SUTs by running benchmarks from the coremark and autobench application suites~\cite{Coremark, Autobench} in the hostile environments produced by our tuning methodology. We show that we can achieve statistical significant slowdowns for $98\%$ of benchmarks. We compare the slowdowns caused by our auto-tuned hostile environments with hand-tuned hostile environments from prior work and show that the slowdowns are comparable and that in some cases we are even able to achieve higher slowdowns.

In summary, our contributions, in order of presentation, are:

\begin{enumerate}
  \item An auto-tuning methodology for hostile environments with the
    aim to cause slowdowns in an SUT; this methodology is portable and
    can be used to automatically tune hostile environments for
    different chips (Section~\ref{sec:creating_a_hostile_environment}).
  \item An illustration of our methodology on five different chips for
    three interference paths: cache, memory bus, and main memory
    systems; we evaluate several natural search strategies and
    slowdown metrics (Section~\ref{sec:experimental_setup}).
  \item An assessment of the extent that our tuned hostile
    environments slowdown the coremark and autobench real-time application suites; we show that in many cases our auto-tuned hostile
    environments are as effective as hand-optimised assembly environments of prior work~\cite{Radojkovic:2012:EIS:2086696.2086713}, and sometimes better (Section~\ref{sec:results}).
\end{enumerate}

The source code for our framework can be found online.\footnote{https://github.com/mc-imperial/multicore-test-harness}

%\begin{itemize}
%  \item We analyse three search startegies and decide which one is the best for finding the parameters of enemy processes.
%  \item We discuss the interference patters that are similar and different across the chips we evaluate and evaluate which chips are most vulnerable to interference.
%  \item We compare our observed interference with that obtained through fine-tuning an enemy process using assembly code.
%\end{itemize}

%\TSComment{Talk to Dan about having this polished to send off for comments}

%\TSComment{make sure to add section numbers when things calm down, especially in the text of contributions}

%\TSComment{Metrics for coremark and autobench: what \% of applications were we able to observe statistically significant slowdowns? Is this similar to prior work?}

\section{Creating a Hostile Environment}
\label{sec:creating_a_hostile_environment}

We now describe in detail our methodology for creating a hostile
environment, which aims to be effective at causing slowdowns in
an SUT through shared resource interference.  After a high-level overview of our
approach (Section~\ref{sec:overview}), we detail the interference paths (shared resources) targeted in this work along with the associated per-resource victim programs (Section~\ref{sec:shared_resources}).  We then explain how we use auto-tuning to search for enemy template parameters per resource (Section~\ref{sec:tuningpertest}) and how hostile environments are constructed from tuned enemy processes (Section~\ref{sec:tuninghostile}). Because different search strategies can be used in the tuning phase, we outline several natural choices (Section~\ref{sec:searchstrategies}).  We conclude the section by describing the care we have taken to ensure validity of the measurements that form the basis of the tuning process (Section~\ref{sec:meas_validity}).

\subsection{Overview of our approach}\label{sec:overview}

The first step in our method consists of identifying possible interference paths, i.e.\ shared resources for which multi-core contention might cause slowdowns. For each one of these paths, we create: (1)~a parameterised \emph{enemy template} that will run in an infinite loop and stress the resource and (2)~a \emph{victim program} that performs a fixed amount of synthetic work, making heavy use of the resource. We tune the parameters of the enemy template using the slowdown it can cause on its corresponding victim program as the objective function. Because the victim program is vulnerable to interference on the target resource, the degree to which it is slowed down serves as a proxy for measuring interference on the associated interference path.

% We use a strategy to pick a Pareto optimal configuration of enemies across cores. The resulting trained enemy can then be applied to a SUT.

Each enemy process is tuned to provoke interference to a \emph{specific} resource. However, we want to develop a hostile environment that is effective at slowing down an arbitrary SUT. A black-box SUT is likely to use multiple resources in complex ways. Thus, we aim to find a combination of tuned enemy processes to be effective across all victim programs. The second step in our methodology involves searching for a combination of tuned enemy processes, with one enemy process running on each non-SUT core of the processor, that is effective in causing interference with respect to multiple resources.

\begin{table}
\small
\centering
\caption{The available parameters of the tuning framework}
\begin{tabular}{p{0.4\linewidth}p{0.5\linewidth}}
\toprule
Parameter & Discussed in this work \\
\midrule
Victim program resource & cache; bus; main memory \\
Enemy template resource & cache; bus; main memory \\
\multirow{1}{*}{Search strategy} & random; sim. ann.; Bayesian opt. \\
Metric & median; max; quantile \\
\bottomrule
\end{tabular}
%\TSComment{is bayesian capitalised? Can you check on that and apply it everywhere? I also think you could get everything on one line here. for example, abbreviate to bayesian opt. and sim. annealing. Don't forget a ; after simulated annealing. This table I think is underselling the generality of the idea; we evaluate these options, but people are also free to plug in whatever they'd want here. I guess I would change Option column to ``discussed in this work'', or something like that. And also, the framework is what exactly? In my mind the framework is something that tunes a hostile environment. In that sense, it does not take an SUT, just the stressed resource (and an associated enemy template and victim program). It is kind of confusing as is. }}
\label{table:framework_parameters}
\end{table}

%\TSComment{Have a look at how I use the booktabs package. I also prefer shading over hlines for different sections} \TSComment{Is there something else to put in this table? There,s a lot of white, space. Perhaps the options could be compressed into a single row}

Table~\ref{table:framework_parameters} summarises the parameters that can be used to configure our auto-tuning framework. In the tuning phase of our methodology, different search strategies can be employed. We evaluate three natural choices: random search, simulated annealing and Bayesian optimisation (see Section~\ref{sec:searchstrategies} for more detailed descriptions of each), and compare their proficiency in finding effective parameters.
%for each of the shared resources.
Since measurements in this domain are noisy, we time multiple runs and analyse the results to best approximate the actual interference. These metrics, along with the steps we have taken to ensure measurement validity, are explained in more detail in Section~\ref{sec:meas_validity}.

%\TSComment{I think here needs to a go a table of the parameters of the autotuning framework, these are: (1) the SUT - this can be a litmus test if the aim is to be portable across applications, or a specific application if an application specific stress is wanted. (2) the search strategy, fuzzing, SA, HC, etc. If the solution space has is structured in a way that can be searched, then these can be used. (3) Metric to compare: average, medium, max, or quantile. (4) related to 3, a statistical confidence in the metric, i.e. the confidence interval}
%\DIComment{This comment was in the intro but I moved it here because I think that the intro is already too cluttered.}

\subsection{Shared Resources and Victim Programs}
\label{sec:shared_resources}

We now discuss several shared resources that can lead to interference between independent applications running in parallel on separate cores. For each type of interference, we first outline the ANSI-C victim program we have designed to be vulnerable to this interference, and then describe a parameterized enemy template which aims to effectively provoke interference through this resource.

\paragraph{Bus}

Buses are used to transfer data between memory and processing elements. To reduce the area of a processor, the bus is often shared between multiple processing cores, requiring some form of arbitration mechanism. There are three main classes of resource arbitration mechanisms: (1)~\textit{time-driven} arbitration uses a predefined bus schedule that assigns time slots to contending components; (2)~\textit{event-driven} arbitration resolves contention at runtime, e.g.\ via a round-robin or first-in-first-out strategy; and (3) a \textit{hybrid} approach that uses different runtime policies for each time slot~\cite{10.1007/978-3-642-40184-8_3}.

% \ADComment{A few sentences on why buses provide an interference path.  Avoid stating the obvious, but do keep in less well-known points (e.g.\ arbitration mechanisms).  Basically think: ``would Ally know this'' and if yes, don't mention it.}

\ourparagraph{Victim Program} The bus victim program reads a series of numbers from a main memory data buffer and increments their value. The increment operation forces the enemy process to bring the numbers from main memory to the CPU. Afterwards, the process writes the numbers back to a second buffer in main memory. The buffers are allocated using \texttt{malloc} and are sufficiently large not to fit in the cache. This entire process ensures that the bus will be kept busy with transfers between main memory and registers.

%\ADComment{Explain the point of incrementing numbers---why not just read and write back?  Justify why this will reach the bus - why won't this just stress the cache?  I guess due to the number of transfers that are made?  In general when describing litmus tests and enemies, keep in mind the resource of interest and concisely explain exactly why the test or enemy should be vulnerable to or good for provoking that interference type.}

%\DIComment{Do we need to mention here that the values are mallocated in main memory?} \ADComment{Yes, good to briefly mention that.}

%\ADComment{A few sentences on the relevant litmus test, explaining why it is intuitively a good fit for buses.  Possibly show code if interesting (definitely not space for all code for all litmus tests, I think).}

\ourparagraph{Enemy Template} We have designed the enemy with the aim of hindering data transfers between the CPU and main memory. It achieves this by performing the same operation as the victim process and thus competes for the same bus resource. Each enemy template has distinct buffers to simplify the design and avoid using any synchronisation mechanism.

The configurable parameters are:
\begin{itemize}
  \item Size of main memory data buffers.
  \item Integer data type used for buffers : \texttt{int8\_t}, \texttt{int16\_t}
  \item Which buffer is used for read and which for writes.
\end{itemize}

Intuitively, the larger the size of the data being transferred, the more contention it will cause. However, there might be patterns where fast transfers through the bus followed by pauses will trigger some timing issues as shown in~\cite{7588111}.

\paragraph{Cache}

Caches are used to reduce latency between the processor and main memory. A cache is much smaller than the main memory and, as a result, only a small subset of main memory can be stored in the cache at one time. There are multiple cache levels, and usually the last level cache is shared between processor cores. As an example, the \IntelJoule has two levels of cache, where the level 1 data and instruction caches are core-local, while the level 2 cache is shared between all cores. Once a cache becomes full, stale data is evicted and replaced with newer data using a variety of policies, e.g., first in first out, least recently used, and random replacement~\cite{HennessyPatterson12}. Because these caching policies have a direct effect on memory latency, multi-core timing analysis must take into account potential interference from contending cores.

%\ADComment{A few sentences on why caches provide an interference path.  Avoid stating the obvious, but do keep in less well-known points (e.g.\ replacement policies, cache partitioning).  Basically think: ``would Ally know this'' and if yes, don't mention it.}

%\TSComment{As a general comment, probably best to avoid custom paragraph macros. Low priority, but this probably deserves its own section and each enemy template would be a subsection}
\ourparagraph{Victim Program} The victim program is configured based on the size of the shared cache, its associativity and its line size. The program allocates an array of the size of the last level cache and then reads and writes to the array in a pattern given by the associativity and the line size. The processor optimises access by storing the array in the last level cache.

%\TSComment{I was thinking that the victim programs were uniform across processors. I guess not, and it makes sense. This should be clarified earlier on, but we probably don't have time. Save this comment though}
%\ADComment{A few sentences on the relevant litmus test, explaining why it is intuitively a good fit for caches.  Possibly show code if interesting (definitely not space for all code for all litmus tests, I think).}

\ourparagraph{Enemy Template} The cache enemy template works by striding over a data buffer and performing a sequence of memory operations (reads or writes). The configurable parameters will force it to access the cache in a chaotic manner, thus working against data locality for which caches are designed to optimise. The cache enemy template is configured by the following parameters:
\begin{itemize}
  \item Data buffer size
  \item Stride value
  \item Series of operations; reads, writes (up to five)
\end{itemize}

\paragraph{Main Memory}

Access to main memory is granted by a shared controller. In a single read or write operation, only one bank of memory can be accessed for single-channel memories and simultaneous requests can lead to delayed requests~\cite{6925999}. The interference issues are similar to those of shared caches, however, the runtime impact of \emph{contended} memory accesses are higher.

%\ADComment{A few sentences on why RAM provide an interference path.  Avoid stating the obvious, but do keep in less well-known points (e.g.\ RAM is divided in banks that can be accessed in parallel).  Basically think: ``would Ally know this'' and if yes, don't mention it.}

\ourparagraph{Victim Program} The victim program allocates a data buffer in main memory and writes random values to it. We ensure that the values are actually written in main memory and not just in the cache by having a sufficiently large buffer that does not fit in the cache for any processor.

\ourparagraph{Enemy Template} The goal of the enemy template is to touch as many memory banks as possible. It allocates a large data buffer and repeatedly selects a random, contiguous sub-region of this buffer, of a given fixed size, and {\tt memset}s the sub-region with random values. A randomly-selected byte is chosen and {\tt memset} is used to write this byte across the region. The following parameters configure the enemy process:
\begin{itemize}
  \item The size of the data buffer
  \item The size of the sub-region
\end{itemize}

%\TSComment{Probably not time, but I think it would be cool to have a big table with all the enemy templates, all the parameters and the range of values that they can take. That would essentially describe the tuning space}

\paragraph{Overlap and Extensibility}

There will be a certain level of overlap between enemy processes. The hardware cache policies can make the memory enemy processes first write data to the cache and therefore also act as a cache enemy. Also, we expect that the memory enemy process will stress the system bus by bringing data from the processor to the memory. However, we assume that each enemy process will concentrate its attack on one resource, e.g.\ the bus enemy, in contrast to the memory enemy, will only read data from the same locations in main memory and therefore will demand less work from the memory controller. This way the enemy processes will complement each other.

It is easy to extend the framework and stress different shared resources that we have not included and that might be unique to specific platforms. This extension can be done by adding a pair consisting of a victim program and an enemy template, including the tunable parameters.

% \ADComment{It's great to acknowledge and discuss this overlap!  However, it would be good to also put up a brief defence as to why we hypothesise that the enemies will be complementary.  The argument is easy for the cache enemy since it uses cache size.  What's less clear to me is why we believe the bus and memory enemies will be better at stressing the bus and RAM, respectively.  I.e., why don't we think the bus enemy will be just as good at stressing RAM than the memory enemy?  Some words of clarification here would be great.}

% \ADComment{Briefly make the point that there is overlap.  Also make the point that if there is some new kind of intereference one wishes to investigate, it's pretty straightforward to do this by adding an appropriate litmus test and template.}

\subsection{Tuning enemy processes per Resource}\label{sec:tuningpertest}

%\ADComment{Be really clear that all this is in the context of one processor.  We don't try to tune across different chips.}

For each processor, we tune the parameters of the enemy templates to cause the highest slowdown in their corresponding victim program. In this section, we describe how this process is performed.

In what follows, let $R$ denote the set of resources to be targeted. In our current work, $R = \{ \Bus{}, \Cache{}, \RAM{} \}$. For a resource $r \in R$ let $V_r$ denote the victim program associated with $r$, and $T_r$ the enemy template associated with $r$. For example $V_{\Bus}$ denotes the victim programs associated with the $\Bus$ resource, and $T_{\Cache}$ the enemy template associated with the $\Cache$ resource.

A template $T_r$ takes a set of parameters drawn from a parameter set $P_r$. For a given parameter valuation $p \in P_r$, let $T_r(p)$ denote the concrete enemy process obtained by instantiating $T_r$ with parameters $p$.

For a victim program $V_r$, template $T_r$ and parameter setting $p \in P_r$, let $\slowdown(V_r, T_r(p))$ denote the slowdown associated with (1) executing $V_r$ in isolation on core 0 (with all other cores unoccupied), compared with (2) executing $V_r$ on core 0, in parallel with an instance of $T_r(p)$ on every other available core.

Our aim is to compute:

\[ \argmax{p \in P_r} \; \slowdown(V_r, T_r(p)) \]

Because $P_r$ is too large to search exhaustively, we use a search strategy to approximate the maximum.  These are discussed in Section~\ref{sec:searchstrategies}.

Let $p_r^{\tuned}$ denote the best parameter setting that was found via search using the chosen strategy.  We refer to the set $\resourcetunedenemies{} = \{T_r(p_r^{\tuned})\; \mid r \in R\}$ as the set of \emph{resource-tuned enemies}.

\subsection{Tuning a Hostile Environment}\label{sec:tuninghostile}

The tuning process described in Section~\ref{sec:tuningpertest} considers the same enemy process running on every available core other than that occupied by the SUT.  We aim to devise a deployment of enemy processes that is effective at inducing interference across all resource types since we do not know the resource usage profile of the SUT a priori. We determine the best configuration of enemy processes by using a strategy similar to the one described in~\cite{Sorensen:2016:EER:2980983.2908114}.

We refer to a configuration of multiple possibly distinct enemy processes running on the non-SUT cores as a \emph{hostile environment}.  More formally, for an $n$-core processor where the SUT runs on core 0, a hostile environment is a mapping from $\{1, \dots, n-1\} \rightarrow \resourcetunedenemies{}$.

We now describe our strategy for choosing a suitable hostile environment from the set of $|R|^{n-1}$ possibilities. First, for each resource $r \in R$, we rank every possible hostile environment according to the extent to which they slow down $V_r$, with the environment that induces the largest slowdown ranked first.  Let $\rankedenvironments(r)$ denote this ranking. This set is much smaller than the tuning set, so we can exhausting run these experiments. The most common processor in our case, a 4 core processor, with 3 shared resources would only require 81 evaluations.

We then select a Pareto optimal hostile environment. This is an environment $e$ such that there does not exist an environment $e' \neq e$ such that for all $r \in R$, $e'$ is ranked more highly than $e$ in $\rankedenvironments(r)$. Being Pareto optimal, $e$ may not be unique and in this case we use a tie breaking mechanism. The tie breaking mechanism consists of selecting the environment that is ranked better in all but one of the $\rankedenvironments(r)$.

\subsection{Search Strategies}
\label{sec:searchstrategies}
To estimate the maximum interference caused to the victim program, we need to find effective parameters for the enemy templates given in Section~\ref{sec:shared_resources}. We intuitively expect the search space of enemy process configurations to be discontinuous with respect to interference, e.g.\ due to caches having fixed parameters that are typically powers of two, memory being organised in banks, etc. Therefore we utilise search strategies that do not make any explicit assumption about the convexity of the cost function and do not rely on gradient information. To do so, we evaluate the following candidates:

\paragraph{\ac{RAN}} \ac{RAN} samples different configurations and remembers the best values. This approach has the advantage of being lightweight and providing a baseline for the more complicated techniques.
\paragraph{\ac{SA}} \ac{SA} is a metaheuristic to approximate global optimisation in a large search space. It is often used when the search space is discrete (e.g., all tours that visit a given set of cities). For problems where finding an approximate global optimum is more important than finding a precise local optimum in a fixed amount of time, simulated annealing may be preferable.
\paragraph{\ac{BO}} Having an unknown objective function, the Bayesian strategy is to treat it as a random function and place a prior over it. The prior captures our beliefs about the behaviour of the function. After gathering the function evaluations, which are treated as data, the prior is updated to form the posterior distribution over the objective function. The posterior distribution, in turn, is used to construct an acquisition function that determines what the next query point should be.

\medskip

There are advantages and disadvantages to each one of these strategies. Random search and simulated annealing can quickly determine the next query point. Bayesian optimisation needs time to remodel the objective function and the acquisition function after each new query is made. On the other hand, it is expected that Bayesian optimisation will only sample points that will increase our knowledge of the problem. In general, one would expect to prefer the first strategies in cases where the cost function is cheap to evaluate and \ac{BO} in cases where the cost function is expensive to evaluate. We evaluate the effectiveness of these approaches in Section~\ref{sec:comparing-searches}.

\subsection{Measurement validity}
\label{sec:meas_validity}

% This subsection is written mostlly as a result of the meeting I had with the people from York. I am sure that a lot of people will have similar concerns so I think it is good to have this part in the paper.
% I got a lot inspiration from "Measuring Microarchitectural Details of Multi- and Many-Core Memory Systems Through Microbenchmarking". They have a cool section dedicated to interfering factors. There is a lot of overalap between what they did and what we did to ensure that the measurement results are valid.

A threat to the validity of our approach is that measurement errors and performance fluctuations due to external factors may cause us to wrongly conclude that our test harness is responsible for slowing down an SUT.  Similar to other approaches that make use of enemy processes~\cite{Fang:2015:MMD:2695583.2687356}, we deal with factors related to the hardware, operating system, and compiler. We also make use of a statistical metric, more specifically quantiles, to refine our results.

\begin{figure}[t!]
    \centering
    \includegraphics[width=\linewidth]{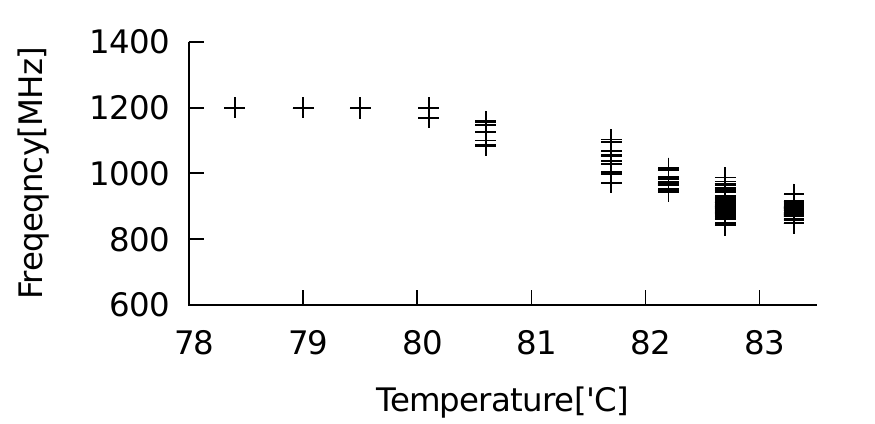}
    \caption{Frequency variation due to temperature on the \PiThree.}
    \label{fig:frequency_variation}
\end{figure}

\paragraph{Hardware} Hardware mechanisms are generally designed to be transparent to the user but can be unpredictable. We took into account the following factors in our design:

\begin{enumerate}

  % NOTE: This is properly implemented
  \item \textit{Frequency throttling due to increased temperature.} When the temperature of a processor increases beyond a limit, frequency throttling can kick in. Figure~\ref{fig:frequency_variation} shows how frequency is affected by temperature on the \PiThree. The data was gathered using a tool called \textit{vcgencmd}. We want to guard against the risk of attributing an SUT slow-down to interference caused by an enemy process when the slow-down is actually due to the raised temperature of the processor.  To mitigate this risk, we measure the temperature at the end of each experiment and discard the result if the temperature has risen above 80 degrees. We empirically found that using this temperature threshold works well on the other devices used.
  % I have this implemented but it is too annoying to run like this. On top of than, I do not think it is that much of a problem.
  % \item \textit{Test harness filling up RAM memory.} The test harness that we designed needs to store measurement results in the main memory. Moreover, more advanced auto-tuning algorithms such as Bayesian Optimisation, use a substantial amount of memory for calculating the next points that need evaluation. However, this can interfere with the execution of the benchmark, effectively acting as another enemy process. For these reasons, we use a different computer to store the measurement results and training data and send it via a network connection to the tested platform.

  \item \textit{Alternating between hot and cold caches.} We flush the cache at the beginning of each experiment as data left from the previous experiment might affect the execution time of the next one.

\end{enumerate}

\paragraph{Operating system}  Modern operating systems are multithreaded and include a range of elements, aimed at efficient execution of a large number of threads. However, this can make thread execution more unpredictable. We use the following techniques to mitigate the effects that the operating system could have on our measurements.

\begin{enumerate}
  \item \textit{Thread migration.} The operating system might decide to migrate the thread to a different core for various reasons. We avoid this by pinning the SUT and the enemy processes on a specific cores using the \texttt{taskset} linux command.
  \item \textit{SUT preemption.} To avoid the kernel from stochastically preempting the SUT, adding the cost of context switching to our measurement time, we run the application at the maximum possible priority.
  % NOTE: I have done this but are not using yet because it forces me to only run the test harness with root access.
  \item \textit{Ensure parallel execution.} The operating system might decide to postpone the startup of any of the enemy processes after the SUT has started, rendering the experiment practically useless. To evaluate the maximum startup latency in different platforms that we considered, we used the latency evaluation framework~\cite{RT_tests} and discovered that the maximum startup latency is generally under 1 ms. To ensure that all the enemy processes are running before the SUT starts executing, we wait 10 ms after all the enemy processes have started, which should be a conservative margin for any evaluation.
  \item \textit{Remove unnecessary software} The interaction between different software can be difficult to predict. To mitigate the chance of this occurring, we removed all software that is not strictly required for our experiments such as any graphical capabilities, logging software and network managers.
\end{enumerate}

%\TSCo0mment{Not enough time for ASPLOS, but it would be cool to have concrete measurements about how each of these issues can screw up timing results}

\paragraph{Compiler} The compiler might optimise away part of the code in the enemy process to reduce execution time, decreasing the stress it is intended to put on specific resources. To avoid this, we generate random numbers at runtime for certain elements, e.g.\ the number written by \textit{memset}. Furthermore, we run the compiler with the \texttt{-O0} flag.

% Why do we need to use statistics
\paragraph{Statistical analysis} We can never truly get rid of all nondeterministic elements of our environment that can interfere with our measurements. For this reason, we need to measure multiple runs and apply a statistical analysis.

% Introducing the idea of quantiles and mentioning that other people are using it too
We need a metric that can reliably estimate the worst-case execution time and ignore the unreliable outliers. Oliveira et. al.~\cite{deOliveira:2013:WYC:2490301.2451140} show how quantiles can be used to compare the latency and end-to-end times of two different Linux schedulers. We follow a similar approach and run the same experiment multiple times and calculate the quantile. Since we are interested in the worst-case behaviour, we would naturally want to select a quantile that is close to the $100^{th}$ one as possible. However, choosing too high of a quantile will not properly disqualify outliers.

\begin{figure}[b]
    \centering
    \includegraphics[width=\linewidth]{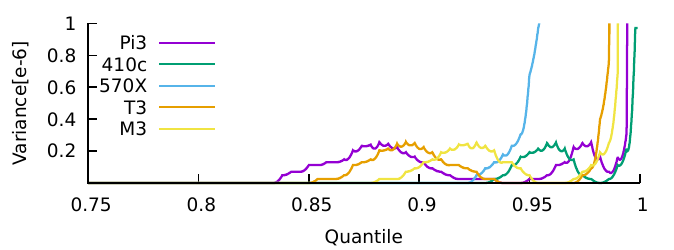}
    \caption{Variation of the quantile.}
    \label{fig:quantile_var}
\end{figure}

% Why we choose the 80th quantile
Figure~\ref{fig:quantile_var} shows the variance of different quantiles for our development platforms. For each board, we ran the coremark benchmark alongside some manually tuned enemy processes and measured the execution time 1000 times. We divided this execution times into 40 sets of 250 data points each and we measured the quantile from the $75^{th}$ to the $100^{th}$. Afterwards we calculated the variance of the 40 sets for each development board and plotted the results. The figure shows how selecting too high a quantile results in noisy data. For that reason, we have chosen the $90^{th}$ quantile as our slowdown metric.

% Introducing the stopping criteria
Another issue that we take into consideration is the number of measurements required for obtaining a reliable value of the quantile. For this reason, we measure 20 times with the same configuration and calculate the $90\%$ confidence interval. If the range of values within the interval is too high, we repeat the process and add more measurements until it decreases to a desired threshold. We do this by calculating the difference between the quantile and the interval endpoints and checking that it is not higher than $5\%$.  However, it often happens that the measurements never converge to the desired threshold. For this reason, we limit the number of measurements to 200.

\section{Experimental setup}
\label{sec:experimental_setup}

We now evaluate the effectiveness of our approach by running it on a collection of embedded development boards, and evaluating its effect on a series of industry benchmarks common in time-constrained domains. In Section~\ref{sec:hard_bench} we present the utilised hardware and afterwards in Section~\ref{sec:comparing-searches} we compare the considered search strategies. We conclude this part with Section~\ref{sec:create_hostile} where we describe how we developed a hostile environment for each platform.

\subsection{Hardware and Benchmarks}\label{sec:hard_bench}

\begin{table}[]
\footnotesize
\centering
\caption{ Benchmarks used to evaluate our approach, along with a short
alias to use in figures.}
\begin{tabular}{p{0.15\linewidth}p{0.5\linewidth}p{0.1\linewidth}}
\toprule
Suite & Benchmark name & Alias \\
\midrule

\multirow{1}{*}{Coremark} & coremark & a \\
\multirow{20}{*}{Autobench} & bitmnp-rspeed-puwmod-4K & b \\
& bitmnp-rspeed-puwmod-4M         &   c   \\
& matrix-tblook-4K                &   d   \\
& matrix-tblook-4M                &   e   \\
& puwmod-rspeed-4K                &   f   \\
& puwmod-rspeed-4M                &   g   \\
& rspeed-idctrn-canrdr-4K         &   h   \\
& rspeed-idctrn-canrdr-4M         &   i   \\
& rspeed-idctrn-iirflt-4K         &   j   \\
& rspeed-idctrn-iirflt-4M         &   k   \\
& ttsprk-a2time-matrix-4K         &   l   \\
& ttsprk-a2time-matrix-4M         &   m   \\
& ttsprk-a2time-pntrch-4K         &   n   \\
& ttsprk-a2time-pntrch-4M         &   o   \\
& ttsprk-a2time-pntrch-aifirf-4K  &   p   \\
& ttsprk-a2time-pntrch-aifirf-4M  &   q   \\
& ttsprk-a2time-pntrch-idctrn-4K  &   r   \\
& ttsprk-a2time-pntrch-idctrn-4M  &   s   \\
& ttsprk-a2time-pntrch-tblook-4K  &   t   \\
& ttsprk-a2time-pntrch-tblook-4M  &   u   \\
\bottomrule
\end{tabular}
\label{table:benchmarks}
\end{table}

\paragraph{Benchmarks} The synthetic victim programs we created are designed to be especially vulnerable to shared resource interference. While these synthetic applications show how we can achieve extreme interference for a specific resource, we are also interested in observing the effects on industry standard benchmarks. These benchmarks are summarised in Table~\ref{table:benchmarks}.

\ourparagraph{EEMBC Coremark}~\cite{Coremark} is a standardised benchmark used for evaluating processors. It is composed of implementations of the following algorithms: list processing (find and sort), matrix manipulation (common matrix operations), state machine (determine if an input stream contains valid numbers), and CRC (cyclic redundancy check).

\ourparagraph{EEMBC Autobench2}~\cite{Autobench} consists of automotive workloads,  including: road speed calculation and finite impulse response filters. This benchmark suite is of interest for the real-time industry and has been used in the evaluation of other works in this domain, e.g.\ \cite{Fernandez:2012:ASN:2380356.2380389, 7588111}.

\paragraph{Hardware} We have chosen a range of development boards, containing both ARM and x86 CPUs to evaluate the portability of our approach. Table~\ref{table:experimental_board} shows the SoC, the architecture and the number of cores for each of them.

\begin{table}[]
\footnotesize
\centering
\caption{ Development boards used to evaluate our approach.  }
%\begin{tabular}{p{0.275\linewidth}p{0.07\linewidth}p{0.145\linewidth}p{0.155\linewidth}p{0.05\linewidth}}
\begin{tabular}{l l l l r}
\toprule
Name & Short name & SoC & Arch & Cores \\
\midrule
Raspberry Pi 3 B & \PiThree & BCM2837 & ARM A53 & 4 \\
DragonBoard 410c & \Dragon & Adreno306 & ARM A53 & 4 \\
Intel Joule & \IntelJoule & 570x & Atom& 4 \\
Nano-PC T3 & \Nano & S5P6818 & ARM A53 & 8 \\
BananaPi M3 & \Banana &A837 & A7 & 8 \\
\bottomrule
\end{tabular}
\label{table:experimental_board}
\end{table}

The operating system can have an impact on the effectiveness of our approach. This impact is minimised to create a fair comparison between the different development boards. We used Debian Linux, as it was available across all platforms.

\subsection{Comparing Search Strategies \label{sec:comparing-searches}}

\begin{table}[]
\small
\centering
\caption{Comparing search strategies when tuning templates against litmus tests. The search strategies are placed in order of effectiveness, with the
ordering symbols described in Section~\ref{sec:comparing-searches}}
\begin{tabular}{p{0.12\linewidth}p{0.22\linewidth}p{0.22\linewidth}p{0.22\linewidth}}
\toprule
Board & Cache & Memory & Bus \\
\midrule
\PiThree & \ac{SA}$<$\ac{BO}$<$\ac{RAN} & \ac{SA}$<$\ac{RAN}$\approx$\ac{BO} & \ac{SA}$<$\ac{BO}$<$\ac{RAN} \\
\Dragon & \ac{SA}$<$\ac{RAN}$<$\ac{BO} & \ac{SA}$<$\ac{RAN}$<$\ac{BO} & \ac{SA}$\approx$\ac{BO}$<$\ac{RAN} \\
\IntelJoule & \ac{SA}$\approx$\ac{RAN}$<$\ac{BO} & \ac{SA}$<$\ac{RAN}$<$\ac{BO} & \ac{SA}$<$\ac{BO}$\approx$\ac{RAN} \\
\Nano & \ac{SA}$<$\ac{RAN}$\approx$\ac{BO} & \ac{SA}$<$\ac{RAN}$<$\ac{BO} & \ac{SA}$\approx$\ac{BO}$\approx$\ac{RAN} \\
\Banana & \ac{SA}$<$\ac{BO}$\approx$\ac{RAN} & \ac{SA}$<$\ac{RAN}$<$\ac{BO} & \ac{SA}$\approx$\ac{BO}$\approx$\ac{RAN} \\

\bottomrule
\end{tabular}

\label{table:search_strategies}
\end{table}

% Explain the methodology in \ref{table:search_strategies}
We now compare the search strategies described in Section~\ref{sec:searchstrategies} and determine which one is the most proficient at finding effective parameters of the enemy templates. We tune the enemy templates using their corresponding victim program as described in Section~\ref{sec:tuningpertest} with all three search strategies tuning for 2 hours. Since all search strategies have a certain degree of randomness and can sometimes get lucky or unlucky (even \ac{BO} starts by randomly sampling its starting points) we perform three runs of each search method for each shared resource. We use the Wilcoxon rank-sum method to test if values from one set are stochastically more likely to be greater than values from another set. This method is non-parametric, i.e.\ it does not assume any distribution of values, and returns a $p$-value indicating the confidence of the result.

% Discuss the results in the table
The results of this experiment can be found in Table~\ref{table:search_strategies} where we constructed an order of the effectiveness of each search method. However, some orders are more confident than others, i.e.\ the ones with a low enough p-value. When the p-value is low (below 0.5) we have higher confidence in the ordering, denoted by the $<$ symbol. However, when the p-value is high (above 0.5) we are not as confident in the ordering, denoted by the $\approx$ symbol. In all cases \ac{SA} seems to perform the worst. \ac{BO} performs well for the memory enemy process and \ac{RAN} for the bus enemy process. However, the difference is not clear for the cache enemy process, with \ac{RAN} and \ac{BO} randomly obtaining the best result.

\ac{RAN} performs well due to the highly irregular search space that the parameters of our enemy templates have. \ac{BO} can intelligently sample points of interest quickly and has reduced chances of getting stuck in a local minimum. It is surprising that \ac{SA} ranks last in our comparison. Most likely the search space is highly irregular, and the algorithm often gets stuck in a local minimum. Of course, \ac{SA} can be configured to focus more on exploration, but then there would be no reason to use it in place of \ac{RAN}.

\subsection{Creating a Hostile Environment\label{sec:create_hostile}}

After determining the appropriate search strategy for each of the enemy processes on each board, we search for the most aggressive parameters that maximise interference. We tune each of the enemy templates and its corresponding victim program with the winning strategies for a more extended period (8 hours).

\begin{table}
\small
\centering
\setlength{\tabcolsep}{10pt}
\caption{ Maximum slowdown obtained using the best search strategy found on the corresponding victim program}
\begin{tabular}{p{0.15\linewidth}p{0.15\linewidth} r r r}
\toprule
\multicolumn{1}{l}{Board} & & Cache & Memory & Bus \\
\midrule

\multirow{2}{*}{\PiThree} & Slowdown & $16.53$ & $6.71$ & $1.38$ \\
& Method & \ac{RAN} & \ac{BO} & \ac{RAN} \\

\rowcolor{Gray} & Slowdown &  $1.81$ & $2.65$ & $1.07$ \\
\rowcolor{Gray}\multirow{-2}{*}{\Dragon } & Method & \ac{RAN} & \ac{BO} & \ac{RAN} \\

\multirow{2}{*}{\IntelJoule} & Slowdown & $1.96$ & $2.65$ & $1.07$ \\
& Method & \ac{BO} & \ac{BO} & \ac{RAN} \\

\rowcolor{Gray} & Slowdown &  $5.29$ &  $1.27$ & $1.17$ \\
\rowcolor{Gray}\multirow{-2}{*}{\Nano } & Method & \ac{BO} & \ac{BO} & \ac{RAN} \\

\multirow{2}{*}{\Banana} & Slowdown & $7.50$ & $49.47$ & $2.18$ \\
& Method & \ac{RAN} & \ac{BO} & \ac{RAN} \\

%\aboverulesepcolor{Gray}
\bottomrule
\end{tabular}
\label{table:maximum_slowdown}
\end{table}

% Describe the results in the \ref{table:maximum_speedup}
Table~\ref{table:maximum_slowdown} presents the maximum slowdown obtained alongside the search strategy used. The most significant slowdowns were obtained for the cache or main memory resources. The bus appears less vulnerable to interference than the other two resources.

% Ramble about processors with same architecture but different implementation
From Table~\ref{table:experimental_board} we see that the \PiThree and the \Dragon have the same architecture, but implemented in different SoCs. The \PiThree is especially vulnerable to cache interference while the \Dragon is much less prone to the same type of interference. It is likely that this can be explained by microarchitectural differences between the the two boards; however, we are not aware of the exact mechanism that causes this difference as low-level details are generally not available for most SoCs on the market today.

\begin{table}[]
\footnotesize
\centering
\caption{ Snippet of rank scores for \IntelJoule. The environment \emph{r} is described by a sequence of 3 letters describing what resource is stressed by each core. For example: CMB indicates that the first core stresses the cache, the second stresses the memory while the third stresses the bus.}
%\begin{tabular}{p{0.04\linewidth}p{0.08\linewidth}p{0.08\linewidth}|p{0.04\linewidth}p{0.08\linewidth}p{0.08\linewidth}|p{0.04\linewidth}p{0.08\linewidth}p{0.08\linewidth}}
\begin{tabular}{p{0.028\linewidth} l r|p{0.028\linewidth} l r| p{0.028\linewidth} l r}
%\toprule
\multicolumn{3}{c|}{Cache} &
\multicolumn{3}{c|}{Main Memory} &
\multicolumn{3}{c}{Bus} \\
%\midrule
\hline
rank & \multicolumn{1}{c}{r} & score & rank & \multicolumn{1}{c}{r} & score & rank & \multicolumn{1}{c}{r} & score \\
1   & MMM & 1.51 & 1   & BBB & 1.41 & 1   & \textbf{MBM} & 1.16 \\
2   & MMB & 1.48 & 2   & \textbf{MBM} & 1.34 & ... & ... & ...   \\
3   & \textbf{MBM} & 1.47 & ... & ... & ...   & ... & ... & ...   \\
... & ... & ...   & ... & ... & ...   & ... & ... & ...   \\
26  & BCC & 1.19 & 26  & BMC & 1.04 & 26  & CBC & 1.02 \\
%\hline
%\bottomrule
\end{tabular}

\label{table:ranked_list}
\end{table}

We now determine the optimal hostile environment for each of the boards using the methodology described in Section~\ref{sec:tuninghostile}. An example of this approach can be seen in Table~\ref{table:ranked_list}, where we have the ranked list of each of the possible environments for the \IntelJoule. For this platform, the \textbf{MBM} configuration is the Pareto optimal, where \textbf{MBM} denotes the hostile environment where the first core stresses the {\bf M}ain memory, the second core stresses the {\bf B}us and the last one also stresses the {\bf M}ain memory.

\section{Results}
\label{sec:results}

%We now present the effectiveness of our approach \TSComment{on a benchmark suite of SUT applications}.
Now we evaluate the effectiveness of the hostile environment on the benchmarks of Section~\ref{sec:hard_bench}. In Section~\ref{sec:evaluate_hostile} we measure how the benchmark runtimes are influenced by our hostile environments and then compare our method with previous approaches in Section~\ref{sec:assembly_stress}.

\subsection{Evaluating the Hostile Environment}\label{sec:evaluate_hostile}

\begin{figure*}[t!] % "[t!]" placement specifier just for this example

\subfloat[\PiThree]{
\includegraphics[width=0.31\linewidth]{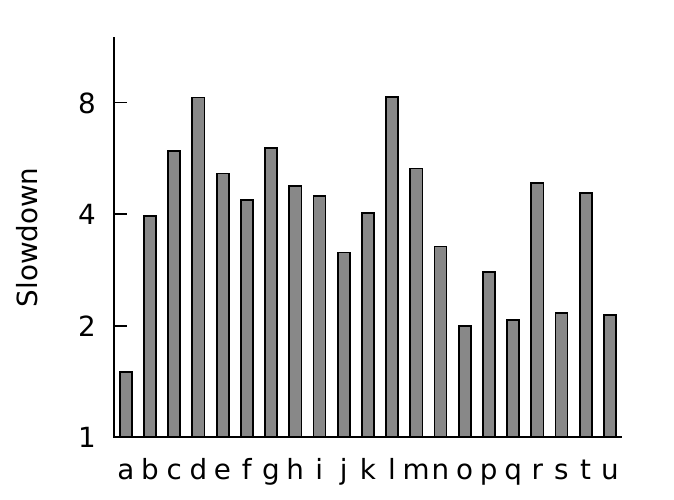}
}%
\subfloat[\Dragon]{
\includegraphics[width=0.31\linewidth]{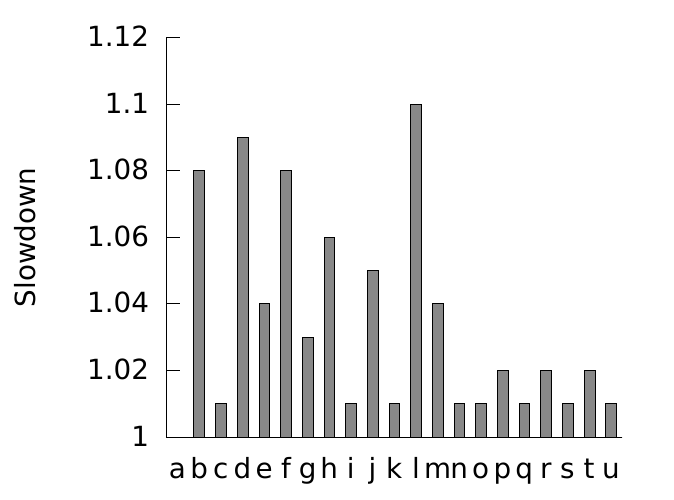}
}%
\subfloat[\IntelJoule]{
\includegraphics[width=0.31\linewidth]{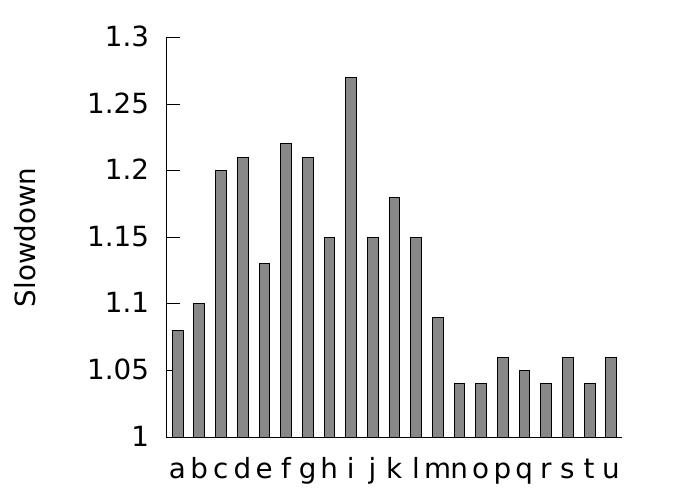}
}\\%
\subfloat[\Nano]{
\includegraphics[width=0.31\linewidth]{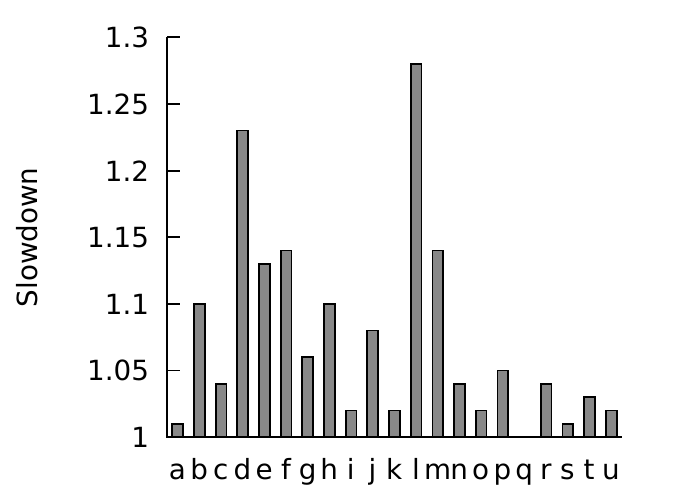}
}%
\subfloat[\Banana]{
\includegraphics[width=0.31\linewidth]{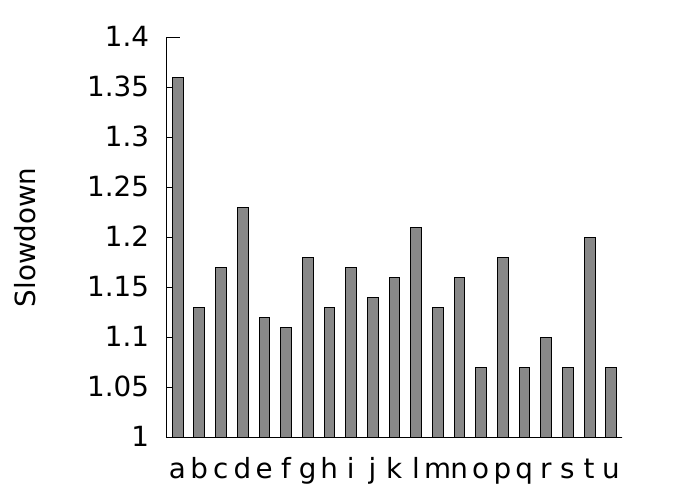}
}%
\subfloat[Geometric mean of slowdown]{
\includegraphics[width=0.31\linewidth]{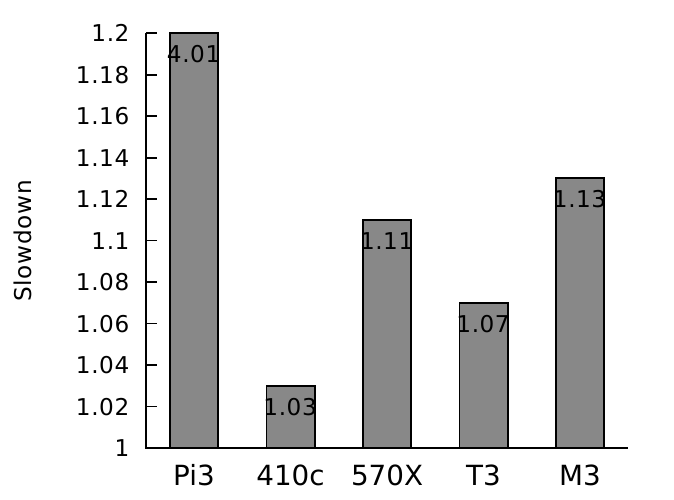}
\label{fig:geometric_mean}
}%
\caption{Slowdowns observed on the available boards. Figures (a)-(e) show the slowdowns obtained for each benchmark on each of the considered development boards. Figure (f) shows the geometric mean of all benchmarks for each board} \label{fig:final_results}
\end{figure*}

With the Pareto optimal hostile environment determined for each SoC, we evaluate its effectiveness on the benchmarks described in Section~\ref{sec:hard_bench}. Figure~\ref{fig:final_results} shows the results of the hostile environment on the benchmarks for each one of the boards. To determine if our slowdowns are statistically significant, we calculated the $90\%$ confidence interval for the benchmarks running in isolation compared to running in the hostile environment. We then proceed to determine if there is any overlap between the two. There are only two cases when they overlap, that is for benchmark \textbf{a} on the \Dragon and for benchmark \textbf{f} on the \Banana. For the considered benchmarks, we can summarise that this approach has a $98\%$ effectiveness of slowing down applications across the benchmark suite we consider.

Figure~\ref{fig:geometric_mean} shows the geometric mean of all benchmark slowdowns for each of the platforms. The \PiThree is the most vulnerable development board in our experiments, while the \Dragon is the most resilient one. This score does not provide a hard guarantee of the timing predictability of any of the tested boards, but it does offer a means by which we can quickly eliminate unreliable platforms. This experiment is in line with the results from Table~\ref{table:maximum_slowdown} where only the victim processes were used, and each resource was stressed individually. More specifically, the large slowdowns in the victim programs can be directly correlated to the large slowdowns in the benchmarks.

\subsection{Comparing with Hand-Written Assembly} \label{sec:assembly_stress}

Previous approaches have relied on hand-crafting assembly enemy processes to assess the maximum slowdown that a given platform can experience. One example of such an approach involves implementing a pointer chasing scenario~\cite{Radojkovic:2012:EIS:2086696.2086713}, in which the enemy process creates a large array of addresses in main memory where each address points to a different location in the same array. The enemy process then starts to navigate the array using assembly code. By utilising assembly code, the compiler is prevented from performing any optimisations. The irregular access pattern contravenes the locality principal needed by the cache to store information efficiently.

\begin{figure*}[]
    \centering
    \includegraphics[width=\linewidth]{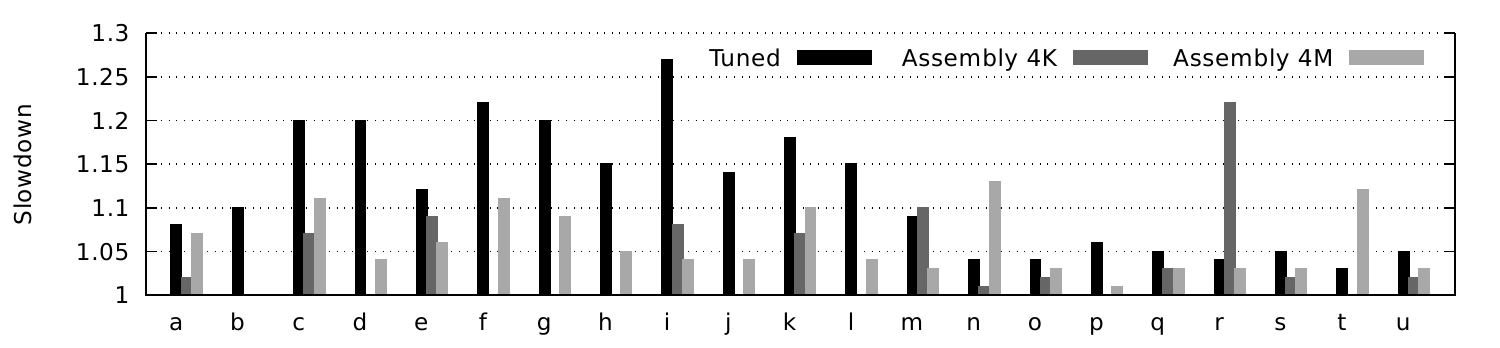}
    \caption{Comparison with two hand-written assembly enemy processes on the \IntelJoule. The first assembly enemy process uses an array of size 4KB to store addresses, while the second one uses an array of size 4MB. }
    \label{fig:assembly_comparison}
\end{figure*}

Using the code provided by the authors of~\cite{Radojkovic:2012:EIS:2086696.2086713}, we evaluated our hostile environment against the previous approach. Since this code is written to assembly language we were only able to execute it on the \IntelJoule, which is the only x86 development at our disposal. We ran the benchmarks alongside the hand-crafted assembly enemy processes and measured the slowdown. Figure~\ref{fig:assembly_comparison} shows the slowdowns observed using the hostile environment and the hand-crafted assembly enemy processes with two different array sizes (4KB and 4MB). We calculated the $90\%$ confidence interval for the obtained results to evaluate if the differences between the two approaches are of statistical significance. The confidence intervals proved to be relatively large, and there was an overlap between the two methods. In 14 cases out of the total 21 benchmarks, our method achieves a higher slowdown. However, out of those, only 11 of them are statistically significant. In the other 7 cases, the assembly approach can reach a higher slowdown, but the confidence intervals always overlap.

Our experiments show a statistically significant higher slowdown in 52\% of the applications, in the remaining 48\% of the cases, there was no statistically significant difference. While our method does not always outperform the hand-written tests, our method has the advantage of being portable, i.e.\ it does not require crafting assembly code for each specific platform.

\section{Related work}

Applications with similar functionality as the enemy processes have been used before in the literature. They have been referred to as \textit{resource stressing benchmarks}~\cite{Radojkovic:2012:EIS:2086696.2086713}, \textit{resource stressing kernels}~\cite{7588111} or \textit{synthetic contenders}~\cite{e49f8f7632ac4b36a20dd2965ea01d1f}. Radojkovic et al.~\cite{Radojkovic:2012:EIS:2086696.2086713} were the first to utilise such techniques by deploying assembly code to measure multi-core interference on real application workloads. They propose a framework for quantifying the maximum slowdown obtained during simultaneous execution by stressing a single shared resource at a time. Their work examines several Intel processors, exploring the extent that the interference from resource stressing benchmarks can slow down real-time software. Nowotsch et al.~\cite{6214768} perform a similar experiment on a multi-core PowerPC-based processor platform and focus specifically on the memory system. The platform allows for different memory configurations and provides several methods for interference mitigation. Regardless of configuration, SUT slowdowns are still observed when executing the resource stressing kernels on distinct cores. Fernandez et al.~\cite{Fernandez:2012:ASN:2380356.2380389} evaluate a multi-core LEON-based processor and run experiments with both a Linux and real-time operating system. Unsurprisingly, the slowdown is mitigated on the real-time operating system, but not eliminated.

Fernandez et al. argue that most resource stressing benchmarks may fail at producing \textit{safe} bounds~\cite{7588111}. Under heavy contention, arbitration policies of shared resources such as round robin and first in first out produce a so-called "synchrony effect" that causes the SUT to suffer a delay that is not as severe as the potential worst-case. They propose a method to improve the bound by varying the injection time between requests to the shared hardware resources. Approaches such as~\cite{7827636, 10.1007/978-3-319-60588-3_7} rely on randomisation of the source code to produce different memory mapping and therefore gather a large set of possible execution times. They utilise a statistical approach called "extreme value theory" and can provide multiple worst-case execution times alongside a confidence factor.

Tuning strategies have been used to optimise different computational aspects, with Ansel et al.~\cite{ansel:pact:2014} showing how such an approach can be used for a variety of domain-specific issues. Wegner et al.~\cite{Wegener1997} use genetic algorithms to find the inputs that cause the longest or shortest execution time. To do so, they formulate the search for such inputs as an optimisation problem. Law et al.~\cite{Law2016} use simulated annealing on a single core processor to maximise code coverage and therefore obtain an estimate of the WCET.

Griffin et al.~\cite{e49f8f7632ac4b36a20dd2965ea01d1f} take a different approach and train a deep linear neural network to learn the relationship between interference and the effect of the SUT execution time. This approach is used to calculate an interference multiplier that can be applied to a previously calculated WCET without interference.

Previous approaches are limited by the need of the developers to tune each resource stressing benchmark for each specific SoC and also can not always detect hidden interference patterns that are specific to the underlying microarchitecture of the system. In contrast, our approach assumes no knowledge of the architecture and microarchitecture of the system and can detect hidden interference patterns automatically.

\section{Conclusions}

We have devised a portable auto-tuning method for determining interference across a wide range of platforms. Our approach is based on configurable enemy processes and does not rely on advanced knowledge of the microarchitectural details of a given platform. For determining the more effective parameters, we compared three different search strategies and determined the better candidates.

We evaluated this method across a wide range of processors, consisting of both ARM and x86 processors using industry standard benchmarks.  Our approach is capable of causing interference in $98\%$ of the cases. We compared the slowdowns caused by our auto-tuned hostile environments with hand-tuned hostile environments from prior work and showed that the slowdowns are comparable, and in some cases, even able to achieve statistically significant higher slowdowns.

\balance

\bibliographystyle{plain}
\bibliography{references}

\end{document}